\documentclass[12pt]{article}
\usepackage{titlesec}
\usepackage{ntheorem}
\titlespacing*{\section}
{0pt} 
{1pt} 
{1pt}  

\titlespacing*{\subsection}
{0pt} 
{1pt} 
{1pt}  

\titlespacing*{\subsubsection}
{0pt} 
{0pt} 
{0pt}  

 \usepackage{pdfpages}
\usepackage{subcaption}
\usepackage{colortbl}
\usepackage{booktabs}
\usepackage{bbm}
\usepackage[normalem]{ulem}

\usepackage{charter}
\usepackage{pgf}
\usepackage{tikz}
\usetikzlibrary{arrows,automata}
\usepackage{booktabs} 
\usepackage[latin1]{inputenc}
\usepackage{verbatim}
\usepackage{subcaption}

\usepackage{xcolor}

\usepackage{booktabs}
\usepackage{bbm}

\usepackage{natbib}

\colorlet{drkblue}{blue!61.8!black}

\usepackage[bookmarks=false]{hyperref}
\hypersetup{
     pdfborder = {0 0 0},
  colorlinks=true,
 citecolor=drkblue,
linkcolor=drkblue,
urlcolor=drkblue,
pdfstartview={XYZ null null 0.80}
}

\usepackage[hyperpageref]{backref}

\usepackage{amsmath}

\usepackage[left=2.5cm,top=3.5cm,bottom=3.5cm,right=3cm]{geometry}

\usetikzlibrary{math}

\usepackage{amssymb,amsmath,xcolor}

\newtheorem{proposition}{Proposition}
\newtheorem{lemma}{Lemma}[section]

\newtheorem{example}{Example}

\theorembodyfont{\rmfamily}
\newenvironment{definition}[1][Definition.]{\noindent\textbf{#1} }

\theorembodyfont{\rmfamily}
\newenvironment{proof}[1][Proof:]{\noindent\textbf{#1} }

\theorembodyfont{\rmfamily}
\newdimen\dummy
\dummy=\oddsidemargin
\newcommand{\MC}{\pi^\rho_W}
\newcommand{\MCw}{W}
\newcommand{\rcf}{\texttt{RCF}}

\newcommand{\be}{\textcolor{blue}{\beta}}
\newcommand{\al}{\textcolor{red}{\alpha}}
\newcommand{\ga}{\textcolor{green}{\gamma}}
\newcommand{\de}{\textcolor{orange}{\delta}}
\newcommand{\te}{\textcolor{blue!61.8!purple}{\theta}}

\newcommand*{\ca}{x}
\newcommand*{\cb}{y}
\newcommand*{\cc}{z}

\newcommand{\Lp}{\pi^\rho_{W^+}}
\newcommand{\pcf}{\texttt{pcf}}
\newcommand{\ccf}{\texttt{ccf}}
\newcommand{\cdf}{\texttt{CDF}}
\newcommand{\M}{\mathcal{M}}
\newcommand{\St}{\mathcal{S}}

\title{Ordered Probabilistic Choice\thanks{We thank the audience of BRIC X and the theory seminar participants at Cornell and Caltech. For this project, Yildiz has received funding from the European Union's Horizon 2020 
research and innovation programme under the Marie Sklodowska-Curie grant agreement no 837702.}}

 \author{\hspace*{0cm} Christopher P. Chambers\thanks{Department of Economics, Georgetown University, ICC 580  37th and O Streets NW, Washington DC 20057. E-mail: \texttt{Christopher.Chambers@georgetown.edu}.} \and Yusufcan Masatlioglu\thanks{University of Maryland, 3147E Tydings Hall, 7343 Preinkert Dr.,  College Park, MD 20742. E-mail: \texttt{yusufcan@umd.edu.}} \and Kemal Yildiz\thanks{Bilkent University, Department of Economics,  Universiteler Mah., Bilkent, 06800, Ankara, Turkey. \ \ \ \ \  E-mail:  \texttt{kemal.yildiz@bilkent.edu.tr.} }}
\date{\today}

\begin{document}

\maketitle

\begin{abstract}
We introduce a novel perspective by linking ordered probabilistic choice to copula theory, a mathematical framework for modeling dependencies in multivariate distributions. Each representation of ordered probabilistic choice behavior can be associated with a copula, enabling the analysis of representations through established results from copula theory. We provide functional forms to describe the ``extremal'' representations of an ordered probabilistic choice behavior\textendash and their distinctive structural properties. The resulting functional forms act as an ``identification method" that uniquely generates heterogeneous choice types and their weights. These results provide valuable tools for analysts to identify micro-level behavioral heterogeneity from macro-level observable data.


   (JEL: D01, D91)

\end{abstract}

       \newpage
                \tableofcontents

\newpage

\section{Introduction}

Ordered probabilistic choice analyzes how individuals make choices among naturally ordered alternatives. Examples abound, including bond ratings, energy consumption, occupational choices, automobile ownership, biological impacts of pesticides, financial investments, and labor force participation rates, among many others. The workhorse model in this area is the random utility model (RUM), which accounts for unobserved heterogeneity in preferences. Although individual choices are deterministic and rational, there is no restriction on conceivable preferences, which makes RUM a highly flexible model. But, this flexibility leads to undesirable identification issues in the form of multiple representations.

The issue of uniqueness presents important conceptual challenges. Theoretically, model identification enables researchers to connect model parameters to observed behavior. Identification ensures that policymakers and mechanism designers can conduct reliable counterfactual analyses. Unique identification ensures that these analyses remain precise. If choice behavior can be represented in multiple ways, counterfactual outcomes may vary across these representations. This becomes particularly critical when evaluating policy questions that depend on the representation.

\cite*{scrum} solves the challenge of non-unique RUM representations by leveraging the inherent order among alternatives. Their work demonstrates that uniqueness is achieved under three assumptions: i) Ordered alternatives, ii) Rational decision-makers, and iii) Ordered individual preference types.  The \textit{single-crossing random utility model} (SCRUM) restricts preference heterogeneity to ordered types while preserving empirical accuracy in capturing individual choices. Expanding on SCRUM, \cite{PRC} explored how ordered types apply to boundedly rational agents, incorporating behavioral factors such as limited attention, willpower, shortlisting constraints, loss aversion, and pro-social behavior. While the resulting \textit{progressive random choice} (PRC) model permits a broader spectrum of agent types, it retains the ordered-type framework to accommodate bounded rationality. PRC allows boundedly rational agents without sacrificing uniqueness.

These findings suggest that it may be the \textit{ordering of types}, rather than \textit{rationality}, that underpins the uniqueness result. This naturally raises two questions: Could alternative, plausible heterogeneity restrictions also yield novel models with unique representations? And given that SCRUM identifies a specific RUM representing the data, what distinguishes this representation from others? What unique properties set SCRUM apart within the class of RUM representations? Our goal is to develop a novel framework to address these fundamental questions by providing a new set of discrete choice models. 

Our contribution is twofold. First, we show that the \textit{ordering of types} itself is not the main driver of the uniqueness result.  We demonstrate this by first establishing that there are wide-ranging models restricting heterogeneity to achieve a unique representation. These models do not require types to be ordered.  To show this, we establish an intriguing connection between ordered probabilistic choice and \emph{copula theory}. This connection provides a robust framework in which each representation of ordered probabilistic choice behavior corresponds to a copula. By leveraging established results from copula theory, we systematically examine the structure of a vast class of representations of ordered probabilistic choice behavior\textendash a task that would otherwise be prohibitively complex.

Second, we provide functional forms to describe the ``extremal'' representations of an ordered probabilistic choice behavior. The resulting functional forms (copulas) act as an ``identification method," that uniquely generates heterogeneous choice types and their weights. These results  provide  valuable tools  for analysts to identify micro-level behavioral choice patterns from macro-level observable  data.

We uncover a class of models that restrict heterogeneity to achieve a unique representation by connecting ordered probabilistic choice and \textit{copula theory}. Copulas in probability theory are initially developed by \cite{sklar1959}, who shows that any joint cumulative distribution over real numbers can be expressed as a copula composed of the marginals' distributions. Thus, a copula isolates the dependence among random variables from the randomness of individual variables.

The key connection is that what we refer to as a \emph{representation} of observed data in discrete choice is closely linked to a copula. A representation is a distribution over choice types, and must be consistent with the observed data, which is a set of probability distributions for each available choice set, such as budgets. We call it a \emph{representation} because the \emph{joint distribution} over choice types must reproduce the observed data as \emph{marginal distributions}. 
Then, it follows from \cite{sklar1959}  that any representation of observed data can be induced via a copula. Additionally, each copula provides a functional form that uniquely determines the heterogeneous choice types and their associated weights from  given probabilistic choice dataset\textendash a process we called an \emph{identification method}.

To explore the implications of this connection, we first observe that  SCRUM\ or PRC   representations (progressive representations) correspond to an extremal copula called \emph{Fr\'echet-Hoeffding upper bound} (the min-copula), which has a particularly tractable form. This connection clarifies the uniqueness of progressive representations while providing us  an explicit functional form to describe the underlying distribution of choice types and their weights, facilitating the identification.   

The connection between the SCRUM and PRC representations  and the \emph{Fr\'echet-Hoeffding upper bound} highlights what sets the progressive representation apart from other RUM representations:
For a given choice type \(t\), the progressive representation assigns a higher total probability to types \textit{dominating} \(t\) compared to any other representation for the same probabilistic data.\footnote{A choice type \textit{ dominates} another if, in all choice sets, it consistently selects an alternative that is the same as or ranked higher than the alternative chosen by the other type according to the reference order.} Consequently, the probability assigned to the choice type maximizing the underlying reference order is maximized under the progressive representation.

The uniqueness result holds as long as the probabilistic model is based on a copula.  One might question if the connection between plausible probabilistic choice models and well-known copulas is merely coincidental.  We suggest that many more connections remain to be discovered.
For illustration, we focus on the \textit{Fr\'echet-Hoeffding} \emph{lower} \textit{bound}. Unlike the upper bound, the lower bound is generally not a copula for more than two marginal distributions, meaning that the lower bound cannot always generate joint distributions from given marginals. Therefore, a model identified by the lower bound inherently possesses empirical content. However, like the upper bound, the corresponding representation is unique when it exists. We uncover the full empirical content of the Fr\'echet-Hoeffding lower bound. 

To demonstrate how copula theory can uncover intriguing probabilistic choice models, we introduce the \textit{1-mistake model}  and show that it is identified by the Fr\'echet-Hoeffding lower bound.
To motivate this model, consider a group of individuals aiming to maximize a common reference order.  Occasionally, they fail to choose the optimal alternative.\@  We term these deviations ``mistakes," which may occur due to cognitive limitations or the use of various decision-making heuristics.\footnote{The  1-mistake model is a special case of the  models examined by \cite{bdds}.} In an \(1\)-mistake model, individuals  are allowed to make a single mistake. This model posits that each  choice type is either entirely rational (free of mistakes) or makes a mistake in a single choice set.   Unlike PRC, the 1-mistake model has empirical content fully captured by a single axiom. Moreover, unlike RUM, it offers an interpretation that avoids imposing unrealistically large behavioral heterogeneity in the population.
In Section 5, we present several other examples of copulas with empirical content that may be useful in applications.

\subsection*{Related literature}
We aim to bridge the fields of ordered probabilistic choice and copula theory, offering a novel perspective on key concepts in discrete choice such as \textit{representation} and \textit{identification}. Ordered probabilistic  choice has a long history, with roots in discrete choice  \citep[e.g,][]{amemiya1981qualitative, small1987discrete,Agresti1984}. Our main objective is to  examine how individuals make probabilistic choices among ordered alternatives. \cite{scrum} has reinvigorated interest in this area  \citep{barseghyan2021discrete,tserenjigmid2021order,turansick2022,yildiz2024,PRC,apesteguia2023random,apesteguia2023testable,petri2023binary,masatlioglu2024growing}. This resurgence in interest stems from the growing availability of detailed choice data and theoretical models.

The literature on copulas in statistics is vast to survey here, but we emphasize that nearly every result we discuss here has a continuous counterpart in this literature.  An excellent textbook treatment is provided by \cite{nelsen2006}; \cite{schweizer2005} is also a standard reference.  The bounds we refer to seem to be named after the contributions by \cite*{frechet1935generalisation,frechet1951tableaux,hoeffding1940}.  Sklar's theorem appears in \cite{sklar1959}. 
The results related to the one-mistake model are understood in statistics as results on negative dependence; fundamental results are due to \cite{dall1972}, see \cite*{lauzier2023} for a modern treatment.  Copulas are extensively utilized in econometrics and finance---\cite{fan2014copulas} provide a systematic treatment. Copulas have been used in other areas of economic theory, for example, in bargaining \cite*{bastianello2019}, in auction \cite*{gresik2011}, in behavioral game theory \cite*{frick2022dispersed}.

\section{Ordered Probabilistic Choice and Copulas }
\subsection{Preliminaries}
 
Let $X$ be a finite set of \textbf{alternatives}. We consider scenarios in which the alternatives possess a natural order. Examples include selecting tax policy according to the total revenue generated, choosing lotteries according to their expected monetary value, determining the number of automobiles owned, choosing the time of day for commuting, comparing insurance offers according to their deductibles, choosing public good provision, and evaluating levels of labor force participation. We call this underlying  order a  \textbf{reference order}, denoted by $\vartriangleright$, which is a complete, transitive, and asymmetric binary relation over $X$.\@  We write $\trianglerighteq$ for
its union with the equality relation.     

Let $N$ be the set $\{1,\dots,n\}$,  whose cardinality is denoted by $n$. Let $\{S_i\}_{i\in N}$ be a family of \textbf{choice sets}, where $S_i\neq \emptyset$. Notably, we do not make the full domain assumption, hence $\{S_i\}_{i\in N}$ could be a strict subset of $2^X$. A (deterministic) \textbf{choice type} $s$ is a list of alternatives $[s_1,s_2,\dots, s_n]$ such that $s_i \in S_i$ for each $i\in N$. Let $\mathcal{S}$ and $\mathcal{S}_R$ be the set of all choice types and all \emph{rational} types, respectively. The reference order $\vartriangleright$ allows us to naturally compare choice types: A choice type $s$ \textbf{dominates} another choice type $s'$ if $s_i$ is $\trianglerighteq$-better than  $s'_i$ for each $i\in N$. With a slight abuse of notation, we also use the notation $\trianglerighteq$ to describe this \textbf{dominance relation} on choice types:  $s\trianglerighteq s'$ if  $s_i \trianglerighteq s'_i$ for each $i\in N$. 

Let $\rho_i(x)$ denote the probability that alternative $x$ is chosen from the choice set $S_i$. Hence $\rho_i$ is a probability distribution over $S_i$. A \textbf{probabilistic choice function} ($\pcf$) $\rho$  is a collection: $\{\rho_i\}_{i \in N}$. Let $\mathcal{P}$ denote the set of all $\pcf$s. Given that our domain is ordered, these $\pcf$s are referred to as \textbf{ordered probabilistic choices}, enabling the definition of a cumulative choice function. The \textbf{cumulative choice function} ($\ccf$) associated to $\rho$ is $P^\rho=\{P^\rho_i\}_{i \in N}$ such that  
$$P_i^\rho(x)=\displaystyle{\sum_{y\in S_i : x\trianglerighteq y} \rho_i(y)}$$ for each $x \in S_i$ and $i\in N$.   
We will use $P$ and $P_i$ instead of $P^\rho$ and  $P^\rho_i$ when the context allows for clarity. Notably, it is the order structure that enables the unique association of a cumulative choice function with a probabilistic choice function. 

As mentioned before, there is a connection between cumulative choice functions and copulas, which are flexible tools for modeling dependence among random variables. A copula creates a multivariate distribution from a given set of random variables \citep{nelsen2006}. Formally,  a \textbf{copula}  is a function \(C: [0,1]^n \to \mathbb{R}\) that satisfies the following three properties.  It is \textit{grounded:} If \(C(u_1, u_2, \ldots, u_n) = 0\) when any \(u_i = 0\). It \textit{has uniform margins:} If \(C(1, \ldots, 1, u_i, 1, \ldots, 1) = u_i\) for any \(i\in N\).  Additionally, \textit{the rectangle inequality} requires \(C\) to induce a nonnegative distribution over any \(n\)-dimensional cube,   $[a_1,b_1]\times[a_2,b_2]\times \dots \times [a_n,b_n]$, ensuring that $C$ is a valid joint distribution.\footnote{This is a variant of the inclusion-exclusion principle (or  $n$-increasing property).\@ The inclusion-exclusion formula ensures the \( C \)-volume of any hyperrectangle is non-negative if \( \mathbf{a} \leq \mathbf{b} \). That is   $\sum_{A\subseteq N}  (-1)^{|A|} C(\mathbf{a}_{A},\mathbf{b}_{N\setminus A})\geq 0$ where \( \mathbf{a}_A \) denotes the coordinates \( a_i \) for \( i \in A \) and  \( \mathbf{b}_{N \setminus A} \) denotes the coordinates \( b_i \) for \( i \notin A \). The alternating signs \( (-1)^{|A|} \) stem from the inclusion-exclusion principle, serving to adjust for overcounting or undercounting in the joint distribution.
  }

\noindent \textbf{Sklar's Theorem:}
\cite{sklar1959}    shows that any joint cumulative distribution over the real numbers can be expressed as a copula composed with the marginal distribution functions of the joint distribution.
Formally, let \( F \) be an \( n \)-dimensional cumulative distribution function ($\cdf$) with marginal distribution functions \( F_1, F_2, \dots, F_n \). Then, there exists a copula \( C \) such that for each \( x_1, x_2, \dots, x_n \in \mathbb{R} \), 
\[
F(x_1, x_2, \dots, x_n) = C(F_1(x_1), F_2(x_2), \dots, F_n(x_n)).
\]


\subsection{Connection to Ordered Probabilistic Choice}

To establish this connection, we first formally define when a probability distribution $\pi \in \Delta(\mathcal{S})$ over choice types represents a given $\pcf$ $\rho$. This requires the choice probability of $x$ from $ S_i$ be equal to the total weights of choice types who choose $x$ from $S_i$.  

\begin{definition}
Let \(\rho\) be an $\pcf$ and \(\pi\in \Delta(\mathcal{S})\) be a probability distribution over choice types. Then, \(\pi\) \textbf{represents}  \(\rho\) if for each $i$ and $x \in S_i$ we have 
$$\rho_i(x)=\sum\limits_{s: s_i=x} \pi(s).$$
\end{definition}

Understanding of probability distributions over choice types that represent $\pcf$s is at the heart of many exercises in probabilistic choice. This motivates us to describe a \textit{choice model} as a set of $\langle \rho,\pi\rangle$ pairs where $\rho$ is the observed data consistent with the model and $\pi$ is an unobservable representation of $\rho$. 

\begin{definition}
A (choice) \textbf{model} $\M$ is a set of $\langle \rho,\pi\rangle$ pairs where $\rho$ is an $\pcf$  and $\pi \in \Delta(\mathcal{S})$ is a probability distribution over choice types that represents $\rho$.
\end{definition}
 We next describe two new objects associated with a model $\mathcal{M}$. For each $\pcf$ $\rho$, let   $I_{\mathcal{M}}(\rho)$ be the set of all representations $\pi$ such that $\langle \rho,\pi\rangle$ is contained in $\M$, i.e., $I_{\mathcal{M}}(\rho):=\{\pi |  \langle \rho,\pi\rangle \in  \M \}$. Note that if there is no representation  $\pi$ such that $\langle \rho,\pi\rangle\in \M$, then $I_{\mathcal{M}}(\rho)=\emptyset$.  Let $\mathcal{P}_{\mathcal{M}}$ be the set of $\pcf$s such that $I_{\mathcal{M}}(\rho)\neq\emptyset$. So, $\mathcal{P}_{\mathcal{M}}$ is the set of $\pcf$s that are \textbf{consistent} with a model $\mathcal{M}$.\footnote{It may be tempting to refer to $\mathcal{P}_{\mathcal{M}}$ as a model. However, two distinct models, $\mathcal{M}$ and $\mathcal{M}'$, can satisfy $\mathcal{P}_{\mathcal{M}} = \mathcal{P}_{\mathcal{M}'}$ even though $\mathcal{M} \neq \mathcal{M}'$ (see footnote \ref{footnote:suleymanov} for an example). This distinction enables a more expressive framework for differentiation.}

A classical representation theorem in decision theory focuses on the properties satisfied by $\mathcal{P}_{\mathcal{M}}$. When $\mathcal{P}_{\mathcal{M}} = \mathcal{P}$, where $\mathcal{P}$ represents the set of all $\pcf$s, the model has no empirical content, as it can account for every conceivable behavior. In contrast, a smaller $\mathcal{P}_{\mathcal{M}}$ enhances the model's predictive power by imposing constraints on the behaviors it can explain, thereby making it empirically meaningful. 
 
 An identification theorem examines the \emph{size of $I_{\mathcal{M}}(\rho)$}, which is the set of representations in $\mathcal{M}$ consistent with the observed behavior $\rho$. A model $\mathcal{M}$ is \textbf{uniquely identified} if $|I_{\mathcal{M}}(\rho)| = 1$ for every $\rho \in \mathcal{P}_{\mathcal{M}}$. This means that for any observed behavior $\rho$ consistent with $\mathcal{M}$, there exists a single admissible representation $\pi$ that explains $\rho$.

 We use the Random Utility Model (RUM) to illustrate the notation introduced above. RUM consists of $\pcf$s that can be represented by a probability distribution over rational types. 
 It is well known that, in general, the distribution over choice types cannot be uniquely identified from probabilistic choice data \citep[e.g.,][]{falmagne, Fishburn1998}, i.e,   $|I_{RUM}(\rho)| \neq 1$ for some $\rho\in \mathcal{P}_{RUM}$.\footnote{\label{footnote:suleymanov}\citet{SULEYMANOV2024} introduces the branch-independent RUM (BI-RUM), a model with the same explanatory power as RUM, i.e., $\mathcal{P}_{RUM} = \mathcal{P}_{BRUM}$. However, BI-RUM can be  uniquely identified: $|I_{BI-RUM}(\rho)| = 1$ for each $\rho \in \mathcal{P}_{RUM}$. Additionally, $I_{BI-RUM}(\rho)\in  I_{RUM}(\rho)$. Thus, while RUM and BI-RUM share the same explanatory power, they differ in their identification properties. Recent research by \cite{caliari2024irrational} introduces a novel model that achieves the same explanatory power as RUM. Critically, however, the authors demonstrate that its identification framework is fundamentally distinct, revealing that the supports of RUM and their proposed representations share no common individual types.}

A key question is: What is the structure of probability distributions over choice types that represent a given $\pcf$? Answering this question is crucial for understanding which population interpretations can be legitimately derived from observed data.

To answer this question, we rely on a \emph{systematic} method of associating a probability distribution over choice types, a representation $\pi$,  to a given $\pcf$ $\rho$. Determining $\pi$ is generally a challenging task, often involving intricate constructions. For certain models, while the existence of such a representation is established, its precise structure remains elusive. This raises the question of whether it is possible to define a functional form\textendash serving as an identification method\textendash that uniquely determines $\pi$ from a given $\pcf$.

\begin{definition}
An \textbf{identification method} for a model $\mathcal{M}$ is a mapping $\mathcal{I}: \mathcal{P}_{\mathcal{M}} \to \Delta(\St)$ such that $\mathcal{I}(\rho) \in I_{\mathcal{M}}(\rho)$ for every $\rho \in \mathcal{P}_{\mathcal{M}}$. 
\end{definition}

If $I_{\mathcal{M}}(\rho)$ contains exactly one representation for every $\rho \in \mathcal{P}_{\mathcal{M}}$, then  $\mathcal{M}$ is  uniquely identified. The connection between copulas and ordered probabilistic choice stems from the observation that copulas serve as concrete examples of identification methods.

We first show that a copula \(C\)  gives birth to a unique probability distribution that represents \(\rho\). Hence, each copula can be seen as an identification method. 
 To see this, for a given copula $C$, we define the mapping $\mathcal{I}^C:\mathcal{P} \to \Delta(\St)$ such that $\mathcal{I}^C(\rho)$ is the probability distribution over types induced by copula $C$ representing $\rho$.  Here, $\mathcal{I}^C(\rho)(s)$  is the weight assigned to the choice type $s\in \St$ such that the following identity holds.
\begin{equation}
\label{ctd}
 \mathcal{I}^C(\rho)(\{s':s \trianglerighteq s'\})=C(P_1^\rho(s_1),\ldots,P_n^\rho(s_n)).
\end{equation}


Identity~\eqref{ctd} defines the probability distribution $\mathcal{I}^C(\rho)$ through its \emph{multivariate} $\cdf$, capturing the distribution over choice types. As in classical statistics, this representation uniquely extends to a probability measure over choice types, thereby representing the given $\pcf$ $\rho$. 
Thus,  each copula has the potential to provide a functional form that uniquely determines heterogeneous choice types and their associated weights from a given probabilistic choice dataset. Therefore, each copula $C$ induces a model $\mathcal{M}^C=\{\langle \rho,\mathcal{I}^C(\rho)\rangle | \rho \in \mathcal{P}\}$, where $\mathcal{I}^C$ is the unique identification method for $\mathcal{M}^C$. 

In section \ref{sec:PRC},  we illustrate that $\mathcal{I}^M$, where $M$ stands for  the min-copula, is the identification method for the progressive random choice model of \cite{PRC}, hence $\mathcal{M}^M=\mathcal{M}_{PRC}$.





\subsection{Fr\'echet-Hoeffding Bounds}

In this subsection, we present a fundamental result in copula theory, which we later employ for our purposes.  \cite{hoeffding1940} and  \cite{frechet1935generalisation, frechet1951tableaux} independently showed that a copula always lies between two specific bounds.

\noindent {{\bf{Theorem:}} 
(Fr\'echet-Hoeffding bounds) For each copula  $C$, $$\max\{\sum^n_{i=1} u_i+1-n,0\}
\leq C(u_1, u_2, \cdots, u_n) \leq  \min\{u_1, u_2, \cdots, u_n\}.$$}
Moreover, these bounds are pointwise sharp, i.e. for each  $u\in [0,1]^n$,
$$\inf_{C} C(u)=\max\{\sum^n_{i=1} u_i+1-n,0\}
\text{ and } \sup_{C} C(u)=\min\{u_1, u_2, \cdots, u_n\}.$$ 

Historically, the \textbf{FH-lower bound}  and the \textbf{FH-upper bound} (\textbf{min-copula}) have been denoted by $W$ and $M$ respectively, and we follow this notation here. While the upper bound is itself always a copula, the lower bound is only a copula in the case of $n=2$.  When \( n = 2 \), these two bounds correspond to distinct forms of extreme dependencies. In the case of the upper bound, the two random variables are perfectly aligned, exhibiting a property known as \textit{comonotonicity}. Conversely, in the case of the lower bound, the two random variables move in opposite directions,  exhibiting a property known as \textit{countermonotonicity}.


Next, we illustrate how these two copulas  
 generate two distinct representations for the same choice data. We consider two disjoint choice sets \(S_1=\{x,y,z\}\) and \(S_2=\{x',y',z'\}\) with marginal choice probabilities \(\rho_1(z) = 0.20\), \(\rho_1(y) = 0.30\), \(\rho_2(z') = 0.40\), and \(\rho_2(y') = 0.35\).  We assume that the reference order is: \(x \vartriangleright y \vartriangleright z\) and \(x' \vartriangleright y' \vartriangleright z'\).
\medskip

  \begin{figure}[ht]
\centering
\begin{subfigure}{.5\textwidth}
  \centering
  \scalebox{.9}{ 
  \includegraphics[width=.8\linewidth]{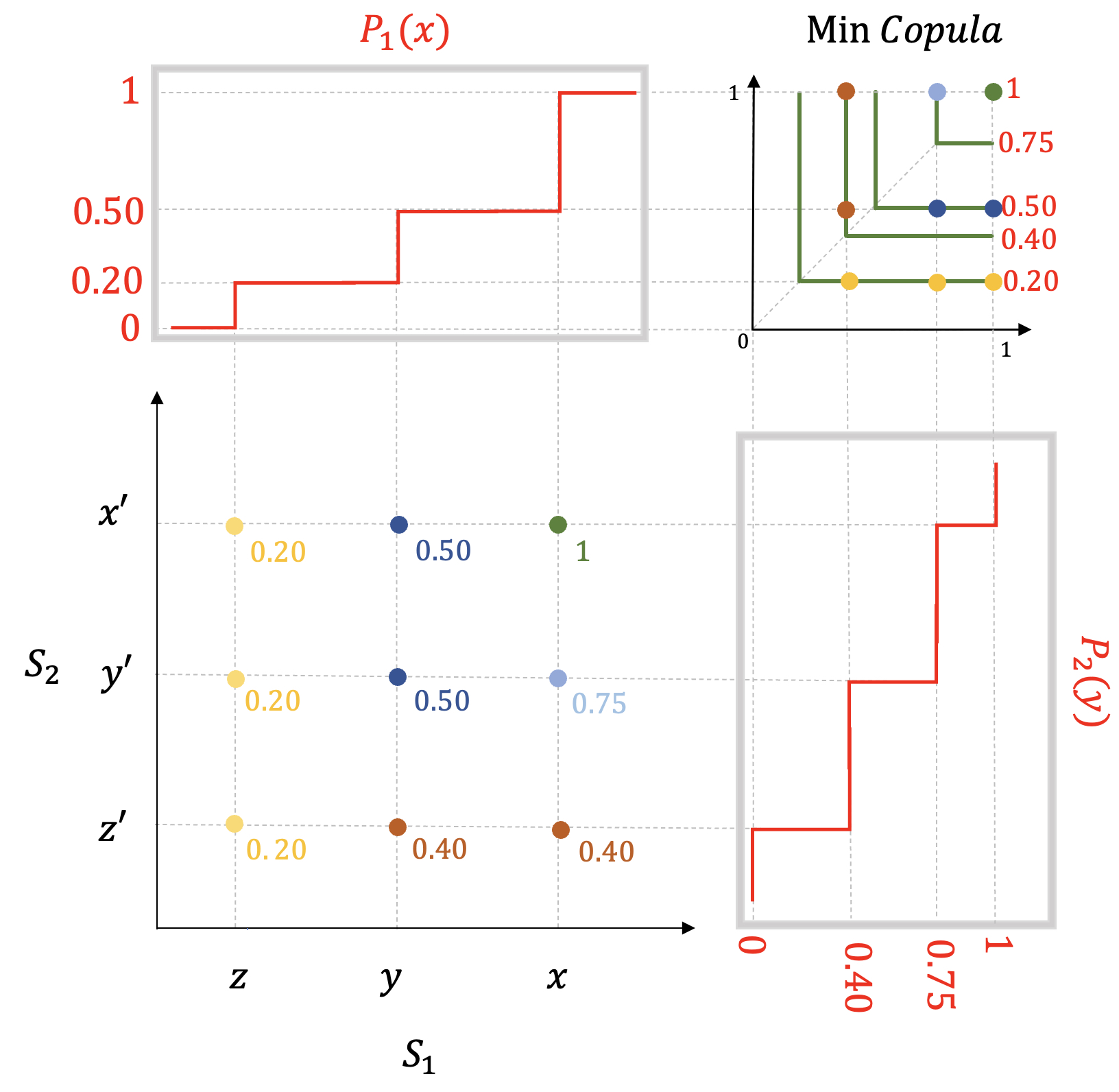}
}
  \caption{\emph{M: The FH-upper bound}}
\end{subfigure}%
\begin{subfigure}{.5\textwidth}
  \centering
  \scalebox{.9}{ 
  \includegraphics[width=.8\linewidth]{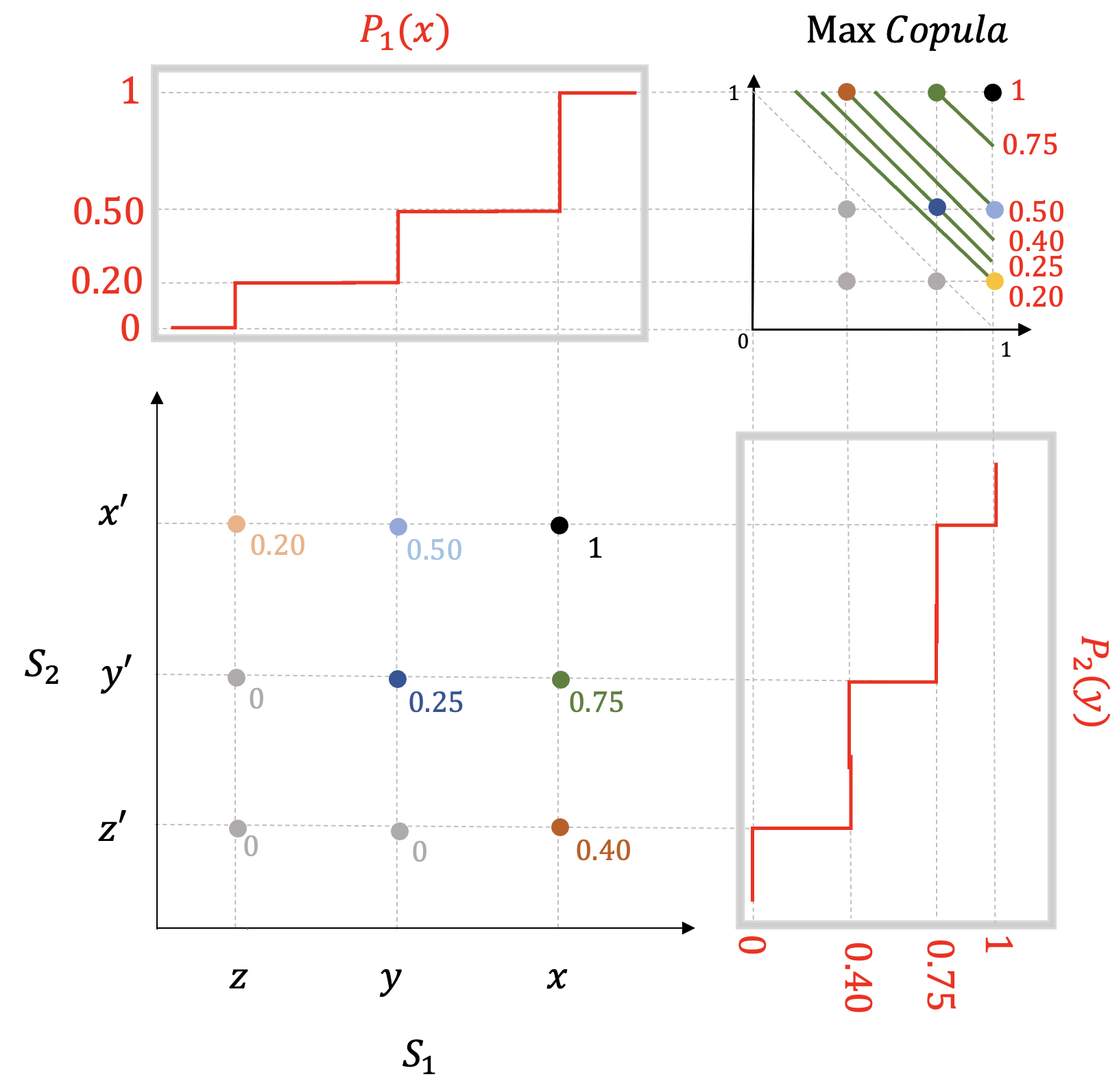}
}
  \caption{\emph{W: The FH-lower bound}}
\end{subfigure}
\caption{\footnotesize{Construction of cumulative distributions over deterministic choice functions for Fr\'echet-Hoeffding bounds involving two choice sets: \(S_1=\{x,y,z\}\) and \(S_2=\{x',y',z'\}\). The marginal choice probabilities are \(\rho_1(z) = 0.20\), \(\rho_1(y) = 0.30\), \(\rho_2(z') = 0.40\), and \(\rho_2(y') = 0.35\). The left panel presents the unique cumulative distribution for \(M\). The cumulative probabilities of types below \([z, z']\) and \([y, y']\) are \(0.20\) and \(0.50\), respectively. The right panel depicts the unique distribution for \(W\). The cumulative probabilities for types below \([z, z']\) and \([y, y']\) are now \(0\) and \(0.25\).}}
\label{fig:copula}
\end{figure}
Figure \ref{fig:copula} illustrates the process of combining marginal distributions  for two separate choice problems into a  joint distribution using the \textbf{FH-lower bound} and \textbf{FH-upper bound}. This approach ensures that the resulting probability structure aligns with the given marginal probabilities and the reference order. Each figure contains two rectangles representing the cumulative marginal distributions for each choice set. The corresponding copula is then applied to determine the probability levels, which are displayed in the top right corner of each figure. These values are subsequently mapped onto the distribution over choice types, as shown in the bottom left corner. The numerical values at each point represent the resulting joint probabilities.

We first calculate the cumulative marginal distributions. We have $P_1(z)=0.20$, $P_1(y)=0.50$, and $P_1(x)=1.00$. Similarly, $P_2(z')=0.40$, $P_2(y')=0.75$, and $P_2(x')=1.00$. Note that $P_1$ and $P_2$ are the same cumulative marginal distributions in both panels since they are based on the same choice data.  In the left panel, we calculate the corresponding cumulative joint distribution using \(M\) (the FH-upper bound). For example, \(M(P_1(y),P_2(z'))=M(0.50,0.40)=0.40\) and \(M(P_1(z),P_2(x'))=M(0.20,1.00)=0.20\). Hence, the cumulative probabilities of the types below \([y, z']\) and \([z, x']\) are \(0.40\) and \(0.20\), respectively. On the other hand, the corresponding cumulative joint distribution using \(W\) (the FH-upper bound) are \(W(P_1(y),P_2(z'))=W(0.50,0.40)=0\) and \(W(P_1(z),P_2(x'))=W(0.20,1.00)=0.20\), displayed on the right panel. Then, the cumulative probabilities of the types below \([y, z']\) are zero. This implies that the probability of the type  \([z, z']\) is also zero.

\begin{figure}[ht]
  \centering
  \includegraphics[width=.7\linewidth]{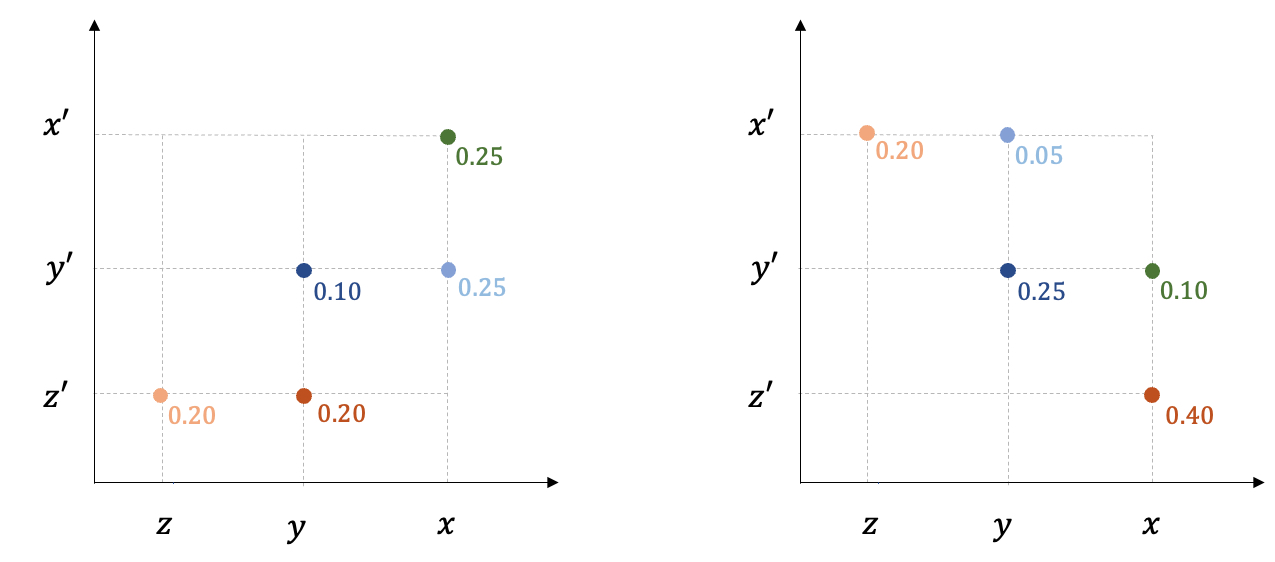}  
\caption{\footnotesize{The representation of the distribution over deterministic choice functions for Fr\'echet-Hoeffding bounds shows two panels. The left panel is based on the upper bound, illustrating the distribution using the min-copula. The right panel depicts the weights induced by the lower bound.}}
\label{fig:copula_rep}
\vspace*{-.5cm}
\end{figure}

Figure \ref{fig:copula_rep} provides the unique weights associated with each type according to \(W\) and \(M\). This is based on the cumulative distributions provided in Figure \ref{fig:copula}. Since \(M(P_1(z),P_2(z'))=M(P_1(z),P_2(x'))=0.20\), 
the representation of \(M\) assigns a probability of \(0.20\) for the type \([z, z']\), while \([z, y']\) and \([z, x']\) have probability of \(0\), which is illustrated on the left panel. Given that the cumulative probability for \([y, z']\) is \(0.40\), we assign a probability of \(0.20\) to \([y, z']\) by subtracting the weight of \([z, z']\). 
\medskip
\begin{table}[ht]
\centering
{\small{\begin{tabular}{l|ccccc}
\toprule
        & \multicolumn{5}{c}{M: The FH-upper bound}             \\
\cline{2-6}
Types   & $[z,z']$ & $[y,z']$ & $[y,y']$ & $[x,y']$ & $[x,x']$  \\
Weights & $0.20$      &  $0.20$        &   $0.10$        &   $0.25$        &  $0.25$          \\
\cline{2-6}
        & \multicolumn{5}{c}{W: The FH-lower bound}             \\
\cline{2-6}
Types   & $[z,x']$ & $[y,x']$ & $[y,y']$ & $[x,y']$ & $[x,z']$  \\
Weights & $0.20$      &  $0.05$        &   $0.25$        &   $0.10$        &  $0.40$ \\
\bottomrule
\end{tabular}}}
\caption{Representations based on the FH-upper bound and the FH-lower  bound.}\label{tab:representation}
\vspace*{-.5cm}
\end{table}

\medskip
Table \ref{tab:representation} presents these two representations. Importantly, the types in the support of $M$ are arranged in a monotonic order, forming a path from $[z, z']$ to $[x, x']$. In contrast, most of the  types in the support of $W$ are not comparable. Additionally, while $M$ assigns a weight of $0.25$ to the highest type $[x, x']$, $W$ assigns it a weight of zero.

\section{Progressive random choice and the FH-upper bound}\label{sec:PRC}

 In this section, we first introduce the\ model of \cite{PRC} (FM) and show that the min-copula $M$ is the identification method for this model. In addition, we show that the connection to copula theory reveals an interesting aspect of their model that was unknown.

FM introduces an ordered probabilistic choice model in which types are ranked based on a fixed characteristic. For instance, consider a set of policies differing in their levels of environmental friendliness. Types are indexed according to their degree of environmental caution. Under this model, a type with a higher index will not choose a less environmentally friendly policy than the one chosen by a lower-indexed type when faced with the same choice problem. This model is called the Progressive Random Choice.

 Formally, a set of distinct choice types $\{s^1, \ldots, s^T\}$ is {\bf{progressive}} with respect to $\vartriangleright$ if  $ \displaystyle{s^t_i \trianglerighteq s^{t+1} _i} $ for each $i\in N$ and  $ t\in \{1,\ldots, T-1\}$.  The progressive structure reduces the heterogeneity of types into a single dimension since choice types gradually become more and more aligned with the choice induced by $\vartriangleright$. For a given reference order $\vartriangleright$, an $\pcf$\ $\rho$ is  a \textbf{progressive random choice} (PRC) if there exists a probability distribution $\pi$ over  $\vartriangleright$-progressive deterministic choice types such that $\pi$ represents $\rho$. Formally,  $$PRC:=\{ \langle \rho,\pi\rangle\ | \pi \text{  has a progressive support and represents } \rho     \}.$$ In their main result, FM shows that every probabilistic choice has a unique PRC representation denoted by $\pi^\rho_{PRC}$, i.e., $\mathcal{P}_{PRC}=\mathcal{P}$ and $I_{PRC}(\rho)=\{\pi^\rho_{PRC}\}$.

We now illustrate that their results follow from the existence of the min-copula. First, note that the min-copula only assigns positive weights to a set of deterministic choice types that are comonotonic, and thus progressive. Furthermore, this representation is always unique.  Therefore, the min-copula is an identification method for PRC. Since the min-copula remains a copula regardless of \(n\), $\mathcal{I}^{M}(\rho)$ is a probability distribution for any \(\pcf\) $\rho$. Hence, $\mathcal{P}_{PRC}=\mathcal{P}$.  This discussion establishes Theorem 1 of FM, highlighting the significant connection between the progressive structure and the min-copula. Additionally, the min-copula provides an explicit functional form for calculating the weights assigned to each deterministic choice function. 

The connection between copula theory and ordered probabilistic choice further reveals another unknown aspect of the PRC representation: For a given choice type \(s\), the PRC representation $\pi_{PRC}^\rho$ assigns a higher probability to choice types dominating \(s\) compared to any other probability distribution $\pi$ over choice types that represents \(\rho\).
It is also true that the  PRC representation of $\rho$ assigns a higher probability to the choice types weakly dominated by \(s\) compared to any representation  $\pi$ of $\rho$.
We next formally state and prove this result.


\begin{proposition}\label{prop1}
Let $\rho$ be an $\pcf$, and let $\pi$ be a probability distribution over choice types that represents $\rho$. Then, for each choice type \(s \in \mathcal{S}\), the PRC representation of $\rho$, denoted by $\pi_{PRC}^\rho$, satisfies the inequalities
\[
\pi_{PRC}^\rho(\{s':s' \trianglerighteq s\}) \geq \pi(\{s':s' \trianglerighteq s\})
\ \ \text{and} \ \ 
\pi_{PRC}^\rho(\{s':s \trianglerighteq s'\}) \geq \pi(\{s':s \trianglerighteq s'\}).
\]
\end{proposition}

\begin{proof}
For the second inequality, since the PRC representation $\pi_{PRC}^\rho$ corresponds to the min-copula, it is sufficient to show that   $\min\{P^\rho_1(s_1),\ldots,P^\rho_n(s_n)\}\geq \pi(\{s':s \trianglerighteq s'\}).$  
To see this, first note that by Sklar's Theorem there exists a copula $C$ such that ${\pi(\{s':s' \trianglerighteq s\})=C(P^\rho_{1}(s_1),\ldots,P^\rho_{n}(s_n))}$. Then, since   min-copula is the FH-upper bound, we have 
$\min\{P^\rho_{1}(s_1),\ldots,P^\rho_{n}(s_n)\} \geq C(P^\rho_{1}(s_1),\ldots,P^\rho_{n}(s_n)).$

To show that the first inequality holds, consider  $\trianglerighteq^{-1}$ which is the inverse of the reference relation $\trianglerighteq$. Let $s, s' \in \mathcal{S}$ be any two choice types that are assigned positive probability by $\mathcal{I}^{M}(\rho)$ under the reference relation  $\trianglerighteq^{-1}$. Since $s$ and $s'$ are comparable under $\trianglerighteq^{-1}$, they are also comparable under $\trianglerighteq$.
Then, since the PRC representation is unique, we must have   $\pi_{PRC}^\rho = \mathcal{I}^{M}(\rho)$ (this equivalence with respect to $\trianglerighteq$ and $\trianglerighteq^{-1}$ that holds for the min-copula  may not hold for other copulas). Therefore, applying the second inequality under $\trianglerighteq^{-1}$ yields
$
\pi_{PRC}^\rho\left(\{s' : s \trianglerighteq^{-1} s'\}\right) \geq \pi\left(\{s' : s \trianglerighteq^{-1} s'\}\right),
$ which establishes the second inequality.
\end{proof}



An immediate implication of Proposition \ref{prop1} is that the probability assigned to the choice type maximizing the underlying reference order is maximized by $\pi_{PRC}^\rho$. Similarly, $\pi_{PRC}^\rho$ maximizes the probability assigned to the choice type minimizing the underlying reference order.
To gain intuition, recall that for each $s,s'$ in the support of $\pi_{PRC}^\rho$, we have  $s\trianglerighteq s'$ or $s'\trianglerighteq s$.  Proposition~\ref{prop1} sharpens this intuition by requiring that, for \emph{every} $s$, the probability of obtaining an $s'$ that is comparable to $s$ (in either direction) under $\trianglerighteq$ is maximized.


As noted in the introduction, SCRUM identifies a specific RUM representation as its unique form. Next, we  formally define SCRUM.   $$SCRUM:=\{\langle \rho,\pi\rangle\ | \ \pi \in \Delta(\mathcal{S}_R)  \text{  has a progressive support and represents}\}.$$ 
The distinction between PRC and SCRUM lies in the support of their representations.  While PRC can accommodate any probabilistic choice, SCRUM imposes an additional restriction where each type must be rational, providing empirical content for SCRUM. This restriction gives SCRUM its predictive power: $\displaystyle{\mathcal{P}_{SCRUM}  \subset \mathcal{P}_{RUM} \subset \mathcal{P}_{PRC}}$. 
Indeed, \cite{scrum} show that choice data satisfies both \textit{centrality} and \textit{regularity} if and only if it has a SCRUM representation.\footnote{To state their axioms, we use their notation where $\rho(x,A)$ represents the choice probability of $x$ in $A$. Then, \textit{Regularity}: If \(x\in B \subset A\), then \(\rho(x, A) \leq \rho(x, B)\). \textit{Centrality}: If \(x \vartriangleright y \vartriangleright z\) and \(\rho(y, \{x, y, z\}) > 0\), then \(\rho(x, \{x, y\}) = \rho(x, \{x, y, z\})\) and \(\rho(z, \{z, y\}) = \rho(z, \{x, y, z\})\).}
This implies that the min-copula assigns positive weights only to the rational types if and only if \textit{centrality} and \textit{regularity} are satisfied. On the other hand, for each $\rho \in \mathcal{P}_{SCRUM}$, $I_{M}(\rho)=I_{PRC}(\rho)=I_{SCRUM}(\rho)$. Thus, the min-copula serves as an identification method for SCRUM. 

With the above observation in hand, we can now examine what sets the SCRUM representation apart from other RUM representations.  Proposition \ref{prop1} provides answers to these questions. SCRUM selects a representation from the set of RUM representations in which the probability assigned to the choice type that maximizes the underlying reference order is maximized.

The min-copula acts as an identification method for any model that incorporates a progressive structure along with certain additional support conditions. In this vein, FM introduces two special cases of PRC: (i) ``less-is-more'' and (ii) ``no-simple-mistakes.'' These special cases, like SCRUM, introduce additional restrictions on choice types. The less-is-more model allows only types that make fewer mistakes with smaller sets. The no-simple-mistakes model ensures that each type does not choose an option in ternary comparisons that has never been chosen in binary comparisons. FM provides behavioral postulates that characterize them. The min-copula can be used to identify the weights in all these models. 

Next, we report another class of models that can be identified using the min-copula.
\cite{yildiz2024} defines a set of choice types $\mathcal{S}^*$ as \textit{self-progressive} if every probabilistic choice function that can be represented as a probability distribution over choice types in  $\mathcal{S}^*$ can also be progressively represented using (possibly different) choice types within $\mathcal{S}^*$. \cite{yildiz2024} characterizes the comprehensive family of self-progressive choice models by establishing that $\mathcal{S}^*$ is self-progressive if and only if $\langle \mathcal{S}^*, \trianglerighteq \rangle$ is a \textit{lattice}: For every pair of choice types in $\mathcal{S}^*$, the choice types corresponding to their \textit{join}---formed by collecting the $\trianglerighteq$-better choices---and their \textit{meet}---formed by collecting the $\trianglerighteq$-worse choices---are also contained in $\mathcal{S}^*$. 
It follows directly that, for any set of types $\mathcal{S}^*$ such that $\langle \mathcal{S}^*, \trianglerighteq \rangle$ is a lattice, the min-copula serves as an identification method for the associated model $\mathcal{M}_{\mathcal{S}^*}:=\{\langle \rho,\pi\rangle\ | \ \pi \in \Delta(\mathcal{S}^*) \text{ that represents } \rho\}.$

   \section{Probabilistic choice induced by the FH-lower bound}\label{sec:minbound}
 We have shown that the min-copula uniquely identifies all special cases of PRC. At first glance, the connection between probabilistic choice and a well-known copula might appear unexpectedly coincidental. However, we now contend that copula theory can serve as a powerful tool for uncovering plausible choice models.
To illustrate this, we first introduce a choice model identified by the FH-lower bound. Then, we explore the full empirical implications of the FH-lower bound.

We first remind you that the FH-lower bound \(W\) is generally not a copula for \(n > 2\). This means that naively applying \(W\) to an $\pcf$ may not result in a probability distribution over choice types. 
  To cover such cases, we leverage the concept of \textit{quasi-copula}  formulated by \cite*{alsina1993} in the bivariate case, and \cite*{nelsen1996} for the general case. Quasi-copulas  generalizes  copulas, relaxing some of their strict requirements while still preserving key properties that make them useful in modeling dependence structures. Namely, the rectangular inequality requirement of a copula is replaced by a Lipschitz condition.\footnote{A function $Q$ satisfies the \textit{Lipschitz condition} if $|Q(\textbf{u})-Q(\textbf{u}')| \leq \sum_i |u_i -u'_i|$ for every $\textbf{u},\textbf{u}'\in[0,1]^n$.}    Now, we define when a quasi-copula identifies an $\pcf$.

\begin{definition}
   A \textbf{quasi-copula} $Q$ \textbf{identifies} an $\pcf$ $\rho$  if  $\mathcal{I}^\rho_Q$ is a probability distribution over $\Delta(\St)$ and represents $\rho$ such that   for each choice type $s$,  
\begin{equation}
\label{fhl}
\mathcal{I}^Q(\rho)(\{s':s \trianglerighteq s'\})=Q(P^\rho_{1}(s_1),\ldots,P^\rho_{n}(s_n)).
\end{equation}
\end{definition}

Identity~\eqref{fhl} holds for any $\pcf$ $\rho$. However,  $\mathcal{I}^Q(\rho)$ might not be a probability distribution. Mirroring the properties of copulas, the corresponding representation is unique whenever it exists. A model induced by a quasi-copula $Q$ is well-defined provided that $\mathcal{I}^Q(\rho)$ forms a valid probability distribution. The set $\mathcal{M}^Q$ consists of all pairs  $\langle \rho,\mathcal{I}^Q(\rho)\rangle $ for which $\mathcal{I}^Q(\rho)$ is a probability distribution. $$\mathcal{M}^Q:=\{\langle \rho,\mathcal{I}^Q(\rho)\rangle\ | \ \mathcal{I}^Q(\rho) \text{ is a probability distribution that represents } \rho \}.$$

Let $\mathcal{P}_{\mathcal{M}^Q}$ denote the set of all $\pcf$s identified by $Q$. Notably, if $Q$ is a copula, then $\mathcal{P}_{\mathcal{M}^Q} = \mathcal{P}$. We interpret any $\pcf$ for which the application of $Q$ fails to yield a probability distribution as being \emph{ruled out} by $\mathcal{M}^Q$ (or, equivalently, by $Q$ itself). Finally, observe that, by construction,  $\mathcal{I}^Q$ uniquely determines $\mathcal{M}^Q$.

\subsection{A model identified by the FH-lower bound}

 As illustrated in the last section, the model induced by $M$ corresponds to PRC. We argue that, with the help of copula theory, we can uncover plausible and intriguing probabilistic choice models. To illustrate this, we first introduce the \(1\)-mistake model, and show that it is identified by the FH-lower bound.   We then investigate the behavioral content of  \(\mathcal{M}_W\).

Consider a set of individuals aiming to maximize their reference order \(\vartriangleright\). However, at times, they may deviate from selecting the best available alternative. We refer to these deviations as ``mistakes" that reflect cognitive limitations or the use of different decision heuristics by individuals. In our \(1\)-mistake model, individuals are allowed to make a single mistake, getting the choice incorrect for at most one choice set. This model posits that each type is either entirely rational (free of mistakes) or makes a mistake in a single-choice set. Formally, a choice type \(s\) is  \textbf{near \(\vartriangleright\)-optimal} if there exists at most one choice set $S_i$ such that \(s_i\) differs from the \(\vartriangleright\)-best element in \(S_i\). Let \(\mathcal{N}_\vartriangleright \) denote all \textbf{near \(\vartriangleright\)-optimal} choice types.

To provide a simple example, let $S=S_1\times S_2 \times S_3$ where  $S_1=\{x,y,z\}$, $S_2=\{x,y\}$, and $S_3=\{x,z\}$. We assume that the reference order is $x \vartriangleright y \vartriangleright z$. Then, for example, $[z,y,x]$ represents the choice function choosing $z, y,$ and $x$ from $S_1, S_2$, and $S_3$.  Each of $[x,x,x]$, $[y,x,x]$, $[z,x,x]$, $[x,x,z]$, and $[x,y,x]$ are nearly $\vartriangleright$-optimal choice types (see Figure \ref{fig:hasse}).

Now, we can define the $1$-mistake model. An $\pcf$  $\rho$ has a \textbf{$1$-mistake} representation with respect to $\vartriangleright$ if there exits a probability distribution $\pi$ over near $\vartriangleright$-optimal choice types that generates $\rho$, i.e., $\textbf{$1$-mistake}:=\{\langle \rho,\pi\rangle\ | \pi \in \Delta(\mathcal{N}_\vartriangleright)  \text{ that represents } \rho \}.$ 

\begin{figure}[ht]
    \begin{center}

  \tikzmath{\x1 = 0; \y1 =0; \x2 = 1.5; \y2 =1.5; \x3 = 4; \y3 =1.5;} 

    \begin{tikzpicture}[ >=latex,glow/.style={%
    preaction={draw,line cap=round,line join=round,
    opacity=0.3,line width=4pt,#1}},glow/.default=yellow,
    transparency group]
            \tikzstyle{every node} = [rectangle]
    \node (aaab) at (\x1,\y1)         {$[\ca, \ca, \ca] $};    
    \node (aacb) at (\x1+\x2,\y1-\y2) {$[\ca, \ca, \cc] $};
    \node (abab) at (\x1-\x2,\y1-\y2) {$[\ca, \cb, \ca] $};
    \node (abcb) at (\x1,\y1-2*\y2)   {$[\ca, \cb, \cc] $};

    \node (baab) at (\x1-\x3,\y1-\y3)         {$[\cb,\ca, \ca] $};
    \node (bacb) at (\x1-\x3+\x2,\y1-\y3-\y2) {$[\cb, \ca, \cc] $};
    \node (bbab) at (\x1-\x3-\x2,\y1-\y3-\y2) {$[\cb, \cb, \ca] $};
    \node (bbcb) at (\x1-\x3,\y1-\y3-2*\y2)   {$[\cb, \cb, \cc] $};

    \node (caab) at (\x1-2*\x3,\y1-2*\y3)         {$[\cc, \ca, \ca] $};    
    \node (cacb) at (\x1-2*\x3+\x2,\y1-2*\y3-\y2) {$[\cc, \ca, \cc] $};
    \node (cbab) at (\x1-2*\x3-\x2,\y1-2*\y3-\y2) {$[\cc, \cb, \ca] $};
    \node (cbcb) at (\x1-2*\x3,\y1-2*\y3-2*\y2)   {$[\cc, \cb, \cc] $};

      
     \foreach \i/\j/\txt/\p in {
      aaab/baab/\text{$\{\ca,\cb,\cc\}$ }/above,
      baab/caab/\text{$\{\ca,\cb,\cc\}$ }/above,
      abab/bbab/\text{ }/above,
      bbab/cbab/\text{ }/above,
      aacb/bacb/\text{ }/above,
      bacb/cacb/\text{ }/above,
      abcb/bbcb/\text{ }/above,
      bbcb/cbcb/\text{ }/above}
            \draw [red,-] (\i) -- node[pos=0.5,sloped,font=\tiny,\p] {\txt} (\j);

         
     \foreach \i/\j/\txt/\p in {
      aaab/aacb/\text{ $\{\ca,\cc\}$}/above,
      abab/abcb/\text{ }/above,
      baab/bacb/\text{ }/above,
      bbab/bbcb/\text{ }/above,
      caab/cacb/\text{ }/above,
      cbab/cbcb/\text{ }/above}
            \draw [blue,-] (\i) -- node[pos=0.5,sloped,font=\tiny,\p] {\txt} (\j);

         
     \foreach \i/\j/\txt/\p in {
      aaab/abab/\text{$\{\ca,\cb\}$ }/below,
      aacb/abcb/\text{ }/above,
      baab/bbab/\text{ }/above,
      bacb/bbcb/\text{ }/above,
      caab/cbab/\text{ }/above,
      cacb/cbcb/\text{ }/above}
            \draw [gray,-] (\i) -- node[pos=0.5,sloped,font=\tiny,\p] {\txt} (\j);

\draw[fill=lightgray,opacity=.1] (abab)   ellipse (1cm and .5cm);
\node at (\x1-\x3-\x2,\y1-\y3-\y2-.3)        {\tiny{ }}; 
\draw[fill=lightgray,opacity=.1] (caab)   ellipse (1cm and .5cm);
\node at (\x1-\x2,\y1-\y2-.3)        {\tiny{ }}; 
\draw[fill=lightgray,opacity=.1] (baab)   ellipse (1cm and .5cm);
\node at (\x1-\x3,\y1-\y3-.3)        {\tiny{ }}; 
\draw[fill=lightgray,opacity=.1] (aaab)   ellipse (1cm and .5cm);
\node at (\x1,\y1-.3)        {\tiny{ }}; 
\draw[fill=lightgray,opacity=.1] (aacb)   ellipse (1cm and .5cm);


           \end{tikzpicture}    \end{center}
          \caption{ \footnotesize This is a Hasse diagram of choice types by the dominance relation according to the reference order-- $\ca \vartriangleright \cb \vartriangleright \cc $.  There are 12 nodes representing all choice types where the first, second, and third elements are the choice from $\{\ca,\cb,\cc\}$, $\{\ca,\cb\}$, and  $\{\ca,\cc\}$, respectively.  
           Two choice types differing only in one choice set are connected by an edge. Different colors of edges represent different choice sets. For example, any red edge indicates a change in the choice in $\{\ca,\cb,\cc\}$. All near $\vartriangleright$-optimal types are by highlighted in gray oval shapes.}
          \label{fig:hasse}
\end{figure}

 Unlike PRC, the 1-mistake model possesses empirical content, i.e., $\mathcal{P}_{\textbf{$1$-mistake}} \neq \mathcal{P}$. In the first part of the following result, we present a postulate that encapsulates the behaviors induced by this model. This postulate asserts that the total probability of mistakes must be less than 1. The second part of the result establishes that the \(1\)-mistake model is identified by the FH-lower bound.

\begin{proposition}\label{thm:p_rnr}
Let $\rho$ be an $\pcf$ and $\bar{s}_i$ be the $\vartriangleright$-best alternative in $S_i$. Then, 
\begin{itemize}
\item[i.] $\rho \in \mathcal{P}_{\textbf{$1$-mistake}}$    if and only if $\sum_{i\in N} (1-\rho_i(\bar{s}_i)) \leq  1$.  
\item[ii.] If $\rho \in \mathcal{P}_{\textbf{$1$-mistake}}$ , then $\rho$ is identified  by the FH-lower  bound.
\end{itemize}
\end{proposition}


\begin{proof} i. If $\rho$  is a 1-mistake model then it immediately follows that  $\sum_{i\in N} (1-\rho_i(\bar{s}_i)) \leq 1$. Conversely, let $\rho$ be an $\rcf$ such that  $\sum_{i\in N} (1-\rho_i(\bar{s}_i)) \leq 1$. Then, for each  $i\in N$ and  $s_i \neq \bar{s}_i$, define the  choice type $s^i$ such that  $s^i_i =s_i$ and $s^i_j = \bar{s}_j$ for each $j\neq i$. Let $\bar{s}$ be the choice type such that  $s_i =\bar{s}_i$ and  for each $ i\in N$. Now, define a distribution $\pi$ over choice types such that   $\pi(s^i)=\rho_i(s_i)$ and $\pi(\bar{s})=1-\sum_{i\in N}(1-\rho_i(\bar{s}_i)) $,  which is  nonnegative  by our assumption.  Thus, $\pi$ generates $\rho$, and has a support consisting of near $\vartriangleright$-optimal  choice types.

ii. Let $ \rho$ be  a $1$-mistake model, and let $s$ be a choice type such that $s_i$ is the alternative chosen from $S_i$ for each $i\in N$. 
Suppose that $s= \bar{s}$. Then,  $\max\{\sum_{i=1}^n P^\rho_i(\bar{s}_i)+1-n,0\}=\max\{1,0\}=1$.  Suppose that there exists unique $i\in N$ such that  $s_i < \bar{s}_i$ .  Then, $\max\{P^\rho_i(s_i)+\sum_{j\neq i}P^\rho_j(\bar{s}_j)+1-n,0\}=\max\{P^\rho_i(s_i),0\}=P^\rho_i(s_i)$. Finally, suppose that there exist at least two  $i,j\in N$ with $s_i < \bar{s}_i$ and $s_j < \bar{s}_j$. Let $\bar{s}_{i}-1$ be the element that is immediately  $\vartriangleright$-worse than $s_i$.   Recall that by Part i,  $\sum_{i=1}^n(1-\rho_i(\bar{s}_i))\leq 1$. Then, we have     $\sum_i P^\rho_i(\bar{s}_{i}-1)\leq 1$.  It follows that  $\sum_{i=1}^n P^\rho_i(s_i)\leq \sum_{i=1}^n P^\rho_i(\bar{s}_{i}-1)+(n-2)\leq 1 + n-2$, as there are at most $(n-2)$ components with $s_i = \bar{s}_i$,  each of which put at most probability $1$ on $\bar{s}_i$.  Thus, $\sum_i P^\rho_i(s_i)+1-n\leq 0$ and $\max\{\sum_i P^\rho_i(s_i)+1-n,0\}=0$.
\end{proof}

The first part of Proposition \ref{thm:p_rnr} states the empirical content of the $1$-mistake model. The second part informs us that the FH-lower bound is an identification method for the $1$-mistake model.

\subsection{The Full Empirical Content of the FH-lower Bound}

We show that our $1$-mistake model is identified by the FH-lower bound. An intriguing question is whether the FH-lower bound can identify additional models beyond the 1-mistake model. To address this, we examine all possible models identified by the FH-lower bound. This analysis allows us to establish a counterpart to the equivalence between progressive random choice and the FH-upper bound for the FH-lower bound. The notion of being  $1$-mistake away from a given choice type is critical for our result.   

In our $1$-mistake model, the mistake-free type was rational---maximizing the reference order--- where mistakes were defined as failures to choose optimally. We now generalize this model by allowing any choice type to serve as the mistake-free type, which we refer to as the core type. Furthermore, mistakes can be described as either downward or upward deviations with respect to the reference order (e.g., over- or under-confidence). For example, suppose the core type is $[z,x,z]$ and mistakes are defined as upward deviations with respect to the order $x \vartriangleright y \vartriangleright z$. Then, all permissible choice types must lie within upward 1-deviation from the core type, yielding the possibilities: $[\cb,\ca,\cc]$, $[\ca,\ca,\cc]$, and  $[\cc,\ca,\ca]$, see Figure \ref{fig:hasse2}. Similarly, $[\cc,\cb,\cc]$ is the only permissible choice type that is downward 1-deviation away from the core type. We are now ready to state our next result. 

\begin{figure}[ht]
    \begin{center}

  \tikzmath{\x1 = 0; \y1 =0; \x2 = 1.5; \y2 =1.5; \x3 = 4; \y3 =1.5;} 

    \begin{tikzpicture}[ >=latex,glow/.style={%
    preaction={draw,line cap=round,line join=round,
    opacity=0.3,line width=4pt,#1}},glow/.default=yellow,
    transparency group]
            \tikzstyle{every node} = [rectangle]
    \node (aaab) at (\x1,\y1)         {$[\ca, \ca, \ca] $};    
    \node (aacb) at (\x1+\x2,\y1-\y2) {$[\ca, \ca, \cc] $};
    \node (abab) at (\x1-\x2,\y1-\y2) {$[\ca, \cb, \ca] $};
    \node (abcb) at (\x1,\y1-2*\y2)   {$[\ca, \cb, \cc] $};

    \node (baab) at (\x1-\x3,\y1-\y3)         {$[\cb,\ca, \ca] $};
    \node (bacb) at (\x1-\x3+\x2,\y1-\y3-\y2) {$[\cb, \ca, \cc] $};
    \node (bbab) at (\x1-\x3-\x2,\y1-\y3-\y2) {$[\cb, \cb, \ca] $};
    \node (bbcb) at (\x1-\x3,\y1-\y3-2*\y2)   {$[\cb, \cb, \cc] $};

    \node (caab) at (\x1-2*\x3,\y1-2*\y3)         {$[\cc, \ca, \ca] $};    
    \node (cacb) at (\x1-2*\x3+\x2,\y1-2*\y3-\y2) {$[\cc, \ca, \cc] $};
    \node (cbab) at (\x1-2*\x3-\x2,\y1-2*\y3-\y2) {$[\cc, \cb, \ca] $};
    \node (cbcb) at (\x1-2*\x3,\y1-2*\y3-2*\y2)   {$[\cc, \cb, \cc] $};

      
     \foreach \i/\j/\txt/\p in {
      aaab/baab/\text{$\{\ca,\cb,\cc\}$ }/above,
      baab/caab/\text{$\{\ca,\cb,\cc\}$ }/above,
      abab/bbab/\text{ }/above,
      bbab/cbab/\text{ }/above,
      aacb/bacb/\text{ }/above,
      bacb/cacb/\text{ }/above,
      abcb/bbcb/\text{ }/above,
      bbcb/cbcb/\text{ }/above}
            \draw [red,-] (\i) -- node[pos=0.5,sloped,font=\tiny,\p] {\txt} (\j);

         
     \foreach \i/\j/\txt/\p in {
      aaab/aacb/\text{ $\{\ca,\cc\}$}/above,
      abab/abcb/\text{ }/above,
      baab/bacb/\text{ }/above,
      bbab/bbcb/\text{ }/above,
      caab/cacb/\text{ }/above,
      cbab/cbcb/\text{ }/above}
            \draw [blue,-] (\i) -- node[pos=0.5,sloped,font=\tiny,\p] {\txt} (\j);

         
     \foreach \i/\j/\txt/\p in {
      aaab/abab/\text{$\{\ca,\cb\}$ }/below,
      aacb/abcb/\text{ }/above,
      baab/bbab/\text{ }/above,
      bacb/bbcb/\text{ }/above,
      caab/cbab/\text{ }/above,
      cacb/cbcb/\text{ }/above}
            \draw [gray,-] (\i) -- node[pos=0.5,sloped,font=\tiny,\p] {\txt} (\j);

\draw[fill=lightgray,opacity=.4] (cacb)   ellipse (1cm and .5cm);
\node at (\x1-\x3-\x2,\y1-\y3-\y2-.3)        {\tiny{ }}; 
\draw[fill=lightgray,opacity=.1] (bacb)   ellipse (1cm and .5cm);
\node at (\x1-\x2,\y1-\y2-.3)        {\tiny{ }}; 
\draw[fill=lightgray,opacity=.1] (aacb)   ellipse (1cm and .5cm);
\node at (\x1-\x3,\y1-\y3-.3)        {\tiny{ }}; 
\draw[fill=lightgray,opacity=.1] (caab)   ellipse (1cm and .5cm);


           \end{tikzpicture}    \end{center}
          \caption{ \footnotesize  Let $\ca \vartriangleright \cb \vartriangleright \cc $ be the underlying order and  $[\cc,\ca,\cc]$ be the core type, highlighted in a darker gray oval shape. Then all upward 1-deviations from the core type ($[\cb,\ca,\cc]$, $[\ca,\ca,\cc]$, and  $[\cc,\ca,\ca]$) are highlighted in a lighter gray oval shape. }
          \label{fig:hasse2}
\end{figure}



Note that our $1$-mistake model is a special case in which every admissible choice type lies within a downward 1-deviation from the rational (reference-order-maximizing) type--no upward deviations are possible (see Figure \ref{fig:hasse}). In contrast, the preceding example allows only upward deviations (Figure \ref{fig:hasse2}). Our next result demonstrates that the most general model identified by the FH lower bound admits deviations in exactly one direction: either all upward or all downward.

In the 1-mistake model, the core type is uniquely pinned down by the reference order. A natural follow-up question is: How can the core type be identified empirically from choice data? The core type can be defined the support of the data. Formally, let  $\rho$ be an $\rcf$. Then, for each $i\in N$, let $S^+_i=\{x\in S_i: \rho_i(x)>0\}$ and $\bar{s}^\rho_i$ ($\underline{s}^\rho_i$) be the $\vartriangleright$-best(worst) alternative in $S^+_i$.\@ Let  $\bar{s}^\rho=[\bar{s}^\rho_1, \ldots, \bar{s}^\rho_n]$ and $\underline{s}^\rho=[\underline{s}^\rho_1, \ldots, \underline{s}^\rho_n]$.  The core type is either $\bar{s}^\rho$ or $\underline{s}^\rho$. If the core type is $\bar{s}^\rho$, then all deviations must be downward as in the $1$-mistake model. If the core type is $\underline{s}^\rho$, then all deviations must be upward. 


\begin{proposition}
\label{IorII}
 An $\pcf$  $\rho$ is identified  by the FH-lower  bound $W$ if and only if 
\begin{itemize}
\item[I.] there exists a   probability distribution over choice types that are  1-mistake away from either  $\bar{s}^p$ or $\underline{s}^p$, or
\item[II.] there exist $i,j\in N$ such that if $k\in N\setminus \{i,j\}$, then    $\rho_k(y)=1$ for some $y\in S_k$. 
\end{itemize}  
\end{proposition}
\begin{proof}
Please see Section \ref{appendix}.
\end{proof}

To interpret this result, consider a population of agents whose choices are centered around a salient choice type \( s^* \), meaning that each type differs from \( s^* \) in at most one choice set. This indicates a relatively homogeneous population. Our characterization reveals that, under mild conditions (specifically when condition II fails to hold), being identifiable by the FH-lower bound necessitates that the salient choice type \( s^* \) be an extreme choice type according to the reference order.

An earlier study \cite{dall1972} establishes the counterpart of our Proposition \ref{IorII} for continuous random variables. In Section \ref{appendix}, we provide a distinct self-contained~proof.

\section{Other examples}

The set of possible copulas is very rich \citep{nelsen2006}. Hence it is beyond the scope of this paper to list them. Instead, we provide several interesting (quasi-)copulas, which could be useful for generating and identifying new models of ordered probabilistic choice. We will use the FH-lower bound ($W$) and FH-upper bound ($M$) in our constructions.

\begin{example}[Independent Copula]\normalfont
 The independent copula, denoted by $\Pi$, represents a structure where each individual variable is independent of the others. It is arguably the simplest copula, defined as    $\Pi(u_1, u_2, \cdots, u_n) :=  \Pi_{i}  \  u_i .$
The independent copula remains unaffected by the reference order and finds applications in diverse fields, including statistical modeling, finance, and machine learning.\end{example}

\begin{example}[Fr\'echet Copula Family] \normalfont \label{exm:Frechet}
The family $\{C_{\alpha}\}$ is a one-parameter collection of quasi-copulas, representing a version of the Fr\'echet copula family \cite{frechet1958remarks}. Each $C_{\alpha}$ is a linear combination of the Fr\'echet-Hoeffding (FH) lower and upper bounds, satisfying $C_0 = W$ and $C_1 = M$:
\[
C_{\alpha}(u_1, u_2, \dots, u_n) := \alpha M(u_1, u_2, \dots, u_n) + (1 - \alpha) W(u_1, u_2, \dots, u_n).
\]
In general, $C_{\alpha}$ is not a copula. However, similar to the FH lower bound $W$, one can derive conditions on the marginal distributions under which $C_{\alpha}$ induces a valid joint distribution. As established in Proposition~\ref{thm:p_rnr}, $\mathcal{I}^W$ yields non-negative weights. Given that $C_{\alpha}$ is a convex combination of $W$ and $M$, the condition required for $\mathcal{I}^{C_{\alpha}}$ is necessarily weaker than that for $\mathcal{I}^W$ and varies monotonically with $\alpha$. Finally, the support of $\mathcal{I}^{C_{\alpha}}$ corresponds to the union of progressive and near-optimal choice types.
 
\end{example}

\begin{example}[Threshold Copula Family]\normalfont
This family is inspired by an example in \cite{nelsen2006}. The threshold copula $C_t$ is constructed as a mixture of the Fr\'echet-Hoeffding lower bound $W$ and upper bound $M$, determined by a threshold level $t$. Specifically, $C_t$ behaves like $M$ when probabilities are below the threshold, and like $W$ when all variables exceed the threshold. Formally,
\[
C_t(u_1, u_2, \dots, u_n) := \begin{cases}
\max\left\{\sum_{i=1}^n u_i + 1 - n,\, t\right\} & \text{if } \forall i\in N,u_i \geq t, \  \\
M(u_1, u_2, \dots, u_n) & \text{otherwise}.
\end{cases}
\]

\noindent Similar to Example~\ref{exm:Frechet}, $C_t$ is a combination of $W$ and $M$, satisfying $C_0 = W$ and $C_1 = M$. However, $C_t$ is not a copula in general, and the support of $\mathcal{I}^{C_t}$ may extend beyond the union of progressive and near-optimal choice types.
\end{example}

New copulas can  be constructed by combining existing ones. A natural way to build multidimensional copulas is to use two dimensional copulas as building blocks, coupling them to form higher-dimensional structures. The following examples illustrate this idea.

\begin{example}[Nested Aggregation]\normalfont
New quasi-copulas can be constructed by grouping choice problems into clusters and applying distinct aggregation functions to model within-group and between-group interactions. For instance, one can define:
\[
C(u_1, u_2, \dots, u_n) := M\big( W\big(W(u_1, u_2), u_3\big), \dots, u_n\big).
\]
This nested structure combines $W$ and $M$ to capture varying dependence patterns across different choice sets.
\end{example}

\begin{example}[Difficulty-Based Grouping]\normalfont
Consider a setting with three groups of choice sets, categorized by difficulty: $G_1 = \{1, \dots, k_1\}$ (easy), $G_2 = \{k_1 + 1, \dots, k_2\}$ (medium), and $G_3 = \{k_2 + 1, \dots, n\}$ (hard).
 A natural way to aggregate these groups is to apply the function $M$ within each group and then $W$ across groups:
\[
C(u_1, u_2, \dots, u_n) := W\big( M(u_1, \dots, u_{k_1}),\ M(u_{k_1+1}, \dots, u_{k_2}),\ M(u_{k_2+1}, \dots, u_n) \big).
\]
Note, however, that $W$ may not necessarily be a copula in this construction.
\end{example}

 

\section{Conclusion}

This paper establishes a novel connection between ordered probabilistic choice and copula theory, providing a new framework for representation and identification in discrete choice models. By leveraging copulas, we show that heterogeneous choice types and their associated weights can be uniquely identified from observed data, offering analysts explicit functional forms for interpretation. Our analysis of the Fr\'echet-Hoeffding bounds reveals the distinctive role of the min-copula in progressive random choice models and the empirical content of the FH-lower bound in capturing bounded rationality, as illustrated by the 1-mistake model. Overall, the copula-based approach opens new avenues for constructing and analyzing rich classes of probabilistic choice models explaining  behavioral heterogeneity.

We believe this paper merely scratches the surface of the vast potential offered by copula theory. We strongly encourage other researchers to explore and expand the frontiers of this promising research area.

\section{Proof of Proposition \ref{IorII}}
\label{appendix}
For each  $s_i\in S_i$, we denote the element that is immediately  $\vartriangleright$-worse (better) than $s_i$   by  $s_i - 1$ ($s_i +1$)
whenever it exists (e.g., for $x\vartriangleright a\vartriangleright b\vartriangleright c\vartriangleright y$, if  $s_i=b$, then  $s_i+1=a, s_i-1=c$).
We denote the set of all choice types by $S$, where $S=\prod_{i\in N}S_i$, and $S_{-j}=\prod_{i\in N\setminus\{j\}}S_i$ for each $j\in N$. Let $\bar{s}$ ($\underline{s} $) be the choice type such that  $s_i =\bar{s}_i$   $(s_i =\underline{s}_i)$ for each $ i\in N$. For each $s,s'\in S$ and $M\subset N$, let $sMs'$ be the element of $S$ that copies $s$ for the components in $M$, and $s'$ for the components in  $N\setminus M$. 

Let $\rho$ be an $\pcf$ that is identified by the  $W$. The next lemma establishes that there is no strictly dominated choice type in the support of    $\pi^\rho_W$, denoted by    $\pi^\rho_{W^+}$. That is, there exist no two choice types $s, s'\in \pi^\rho_{W^+}$ such that $s_i \vartriangleright s'_i$ for each $i\in N$.

\begin{lemma} \label{l_antichain} Let $\rho$ be an $\rcf$ that is identified by    $\MCw$.\@ Then, the set of choice functions that appear in the support of   $\pi^\rho_W$ is an $\rhd$-antichain.\end{lemma}

\begin{proof} By contradiction,  suppose that there exist $s,s' \in \pi^\rho_{W^+}$ such that $s_i \vartriangleright s'_i$ for each  $i\in N$.    Let $F^\rho_W$ be the $\cdf$ associated with
$W$ and $\rho$. That is,    
$F^\rho_W(s)=W(P^\rho_{1}(s_1),\ldots,P^\rho_{n}(s_n))$ for each $s\in S$ (equivalently    $F^\rho_W(s)=\sum_{t: s\trianglerighteq t} \pi^\rho_W(t)$). Now,  let $[s-1]$ be the element of $S$ such that $[s-1]_i=s_i-1$ for each $i\in N$.\footnote{We will use the notation $[s-1]Ms$ for the element of $S$ for which when $i\in M$, $([s-1]Ms)_i=[s-1]$, and otherwise for $i\notin M$, $([s-1]Ms)_i=s$.} Then,\begin{equation}
\label{eq1}
\MC(s)=\sum_{M\subseteq N}   (-1)^{|M|} \ F^\rho_W([s-1]Ms).
\end{equation}
Since $s'$ is in the support of   $\MC$,   for each $t\in S$ such that  $s \trianglerighteq t \trianglerighteq s'$, we have $F_W(t)=\sum_i P^\rho_i(t_i)-n+1>0$.  Note that  $s \trianglerighteq [s-1]Ms \trianglerighteq s'$  for every $M\subseteq N$. It follows that   
\begin{equation}
\label{eq2}
\pi^\rho_W(s)=\sum_{M\subseteq N}(-1)^{|M|}\left(\sum_{i\in M}P^\rho_i(s_i)+\sum_{i\in N\setminus M}P^\rho_i(s_i-1)-n+1\right)
\end{equation}
which can be decomposed into two components each of which equals zero by the principle of inclusion and exclusion. 
\end{proof}

 The following result improves  Lemma \ref{l_antichain} by stating that any two choice functions in   $\pi^\rho_{W^+}$  can differ at most in two choice sets. Moreover, if they deviate at two choice sets, say $S_i$ and $S_j$, then we must have $s_i \vartriangleright s'_i$ and $s'_j \vartriangleright s_j $.

\begin{lemma} \label{p_min2diff} 
If  $s$ and $s'$  are  in the support of $\MC$, then

i.   $|\{i:s_i\neq s'_i\}|\leq 2$, and 

ii. if $|\{i:s_i \neq s'_i\}|=2$, then $s_j \vartriangleright s'_j$ and $s'_k \vartriangleright s_k$ for some $j,k\in N$.
\end{lemma}

\begin{proof} For clarity, we replace $\MC$ with $\pi$ and $F^\rho_W$ with $F$. Then, since $\pi$ generates $\rho$, we can rewrite   $F(s)=1-n+\sum_{i\in N} P^\rho_i(s_i)$  as
\begin{equation}
\label{eq4}
F(s)=1-n+\sum_{i\in N} \sum_{s_i \trianglerighteq y_i}\sum_{z_{-i}\in S_{-i}}\pi(y_i,z_{-i}).
\end{equation}
Next, for each $k\in \{1,\ldots, n\}$, let $S^k=\{t\in S: k=|\{i: s_i \trianglerighteq t_i\}|\}$. Since, by Lemma \ref{l_antichain}, there is no element $t^*\in \pi^\rho_{W^+}$ with $s\rhd t^*$,  we get 
\begin{equation}
\label{eq8}
F(s)=1-n+n \sum_{t\in S^n} \pi(t)+ \sum_{k=1}^{n-1}k\sum_{t\in S^k}\pi(t).
\end{equation}
Since, by definition,   $F(s)=\sum_{t\in S^n} \pi(t)$,  it follows that  
\begin{equation}
\label{eq6}
n-1 = (n-1) \sum_{t\in S^{n}\cup S^{n-1}} \pi(t) + \sum_{k=1}^{n-2}k\sum_{t\in S^k}\pi(t).
\end{equation}
Now, if $\pi(t)>0$ for some  $t\in S^k$ where  $k<n-1$, then this equality fails to hold. Therefore, $\pi(t)>0$ only if $|\{i:s_i \trianglerighteq t_i\}|\geq n-1$.\@
It follows that $|\{i:s_i \trianglerighteq s'_i\}|\geq n-1$. Symmetrically, $|\{i: s'_i \trianglerighteq s_i\}|\geq n-1$.\@  
  Thus,  we conclude that i. and ii. hold.
\end{proof}

Our next lemma provides the last stepping stone to prove Proposition \ref{IorII}. 

\begin{lemma}
\label{IIfails}
 Let $\rho$ be an $\rcf$ that  is identified by  $W$. Suppose that  $\rho$  fails to satisfy part II of Proposition \ref{IorII}. Let     $\al, \be\in \Lp$ such that   $\al_i\vartriangleright \be_i$ and  $\be_j \vartriangleright \al_j$. Then, there exists   $\ga \in \Lp$ and  $k^*\in N$ such that 
\begin{itemize}
\item[1.]   $\ga_{k^*}\neq \al_{k^*}=\be_{k^*}$, and 
\item[2.] If $\al_{k^*}\vartriangleright\ga_{k^*}$ then $\al$ and $\be$ are 1-mistake away from $\bar{s}^p$; if $\al_{k^*}\vartriangleleft\ga_{k^*}$ then $\al$ and $\be$ are 1-mistake away from $\underline{s}^p$.
\end{itemize}  
\end{lemma}
\begin{proof}
 Note that, by Lemma \ref{p_min2diff} part i,
 we have  $\al_k=\be_k$ for each $k\in N\setminus\{i,j\}$.
Since II fails to hold, there exists   $k^*\in N\setminus\{i,j\}$ such that $p_{k^*}(\al_{k^*})<1$. 
It follows that there exists   $\ga \in \Lp$ such that $\ga_{k^*}\neq \al_{k^*}=\be_{k^*}$.\@ Thus, 1 holds. Next, suppose w.l.o.g.\@ that $\displaystyle{ \al_{k^*}\vartriangleright\ga_{k^*}}$.\@ 
Then,  by Lemma \ref{p_min2diff} part ii, for each $s\in  \{\be,\al\}$, it must be that $\ga_i\trianglerighteq s_i$ and $\ga_j\trianglerighteq s_j$.\@ Therefore, $\ga_i\trianglerighteq \al_i\vartriangleright\be_i$ and $\ga_j\trianglerighteq \be_j\vartriangleright\al_j$.\@ Then, by Lemma \ref{p_min2diff} part i, for $\ga$ not to differ from $\al$ and  $\be$ on more than two components, we must have $\ga_i=\al_i$, $\ga_j=\be_j$, and   $\be_{k}=\al_{k}=\ga_{k}$ for each  $k\in N\setminus\{i,j,k^*\}$. 

In what follows, we show that $\al$ and $\be$ are 1-mistake away from $\bar{s}^p$  (had we  supposed that $\displaystyle{\ga_{k^*}\vartriangleleft \be_{k^*}=\al_{k^*}}$, we would be showing that $\al$ and $\be$ are 1-mistake away from $\underline{s}^p$).\@  That is, we claim that $\al_i=\bar{s}^\rho_i$,  $\be_j=\bar{s}^\rho_j$, and   $\bar{s}^\rho_k=\al_k=\be_k$ for each $k\in N\setminus\{i,j\}$.  By contradiction, suppose that this fails to hold for some $l\in N$.      

\noindent \textbf{Case 1:} Suppose that $l\in N\setminus\{i,j\}$. It follows that there exists   $\de \in \Lp$ such that      $\de_l\vartriangleright  \be_l=\al_l$.\@ By Lemma \ref{p_min2diff} part ii, for each $s\in  \{\al,\be\}$, it must be that $\de_i\trianglelefteq s_i$ and $\de_j\trianglelefteq s_j$. It follows that $\de_i\trianglelefteq \be_i\vartriangleleft\al_i\trianglelefteq \ga_i$ and $\de_j\trianglelefteq \al_j\vartriangleleft\be_j\trianglelefteq \ga_j$.
Thus, $\de$ is dominated by $\ga$   on components $i$ and $j$, contradicting to  Lemma \ref{p_min2diff} part ii.

\noindent \textbf{Case 2:} Suppose that $l\in \{i,j\}$. Suppose w.l.o.g.\@ that $l=i$.\@ 
It follows that there exists   $\de \in \Lp$ such that      $\de_i\vartriangleright \al_i$.\@  Since $\al_i=\ga_i$, we already have  $\displaystyle{ \de_{i}\vartriangleright\ga_{i}}$. Next, we show that    $ \de_{k^*}\vartriangleright\ga_{k^*}$,  
  and thus obtain a contradiction to  Lemma \ref{p_min2diff} part ii.
Since $\de_i\vartriangleright \al_i\vartriangleright\be_i$, by Lemma \ref{p_min2diff} part ii, for each $s\in  \{\al,\be\}$, we have  $\de_j\trianglelefteq s_j$ and $\de_k\trianglelefteq s_k$ for each  $k\in N\setminus\{i,j\}$. It follows that  $\de_j\trianglelefteq \al_j\vartriangleleft\be_j$.
Since   $\de_i\vartriangleright\be_i$ and $\de_j\vartriangleleft\be_j$, by  Lemma \ref{p_min2diff} part i,    $\de_{k}=\be_{k}$ for each  $k\in N\setminus\{i,j\}$. Since
 $\displaystyle{\be_{k^*}\vartriangleright \ga_{k^*}}$, it follows that   $ \de_{k^*}\vartriangleright\ga_{k^*}$. \end{proof}

\begin{proof}
[Proof of Proposition \ref{IorII}]
Since the if part is clear, we proceed with the only if part. Suppose that II fails to hold.\@ Then, we show that I holds.  If  each distinct   $\al, \be\in \Lp$ differ on a single component, then II  holds.\@  Since we suppose that this is not the case, let  $\al, \be\in \Lp$ such that    $\al_i\vartriangleright \be_i$ and  $\be_j \vartriangleright \al_j$.
Then, let $\ga$ and $k^*$ be as described in  Lemma \ref{IIfails}.  Suppose w.l.o.g that $\ga_{k^*}\vartriangleright\al_{k^*}$, and thus   $\al$ and $\be$ are both 1-mistake away from $\bar{s}^p$. It follows that $\al_i=\bar{s}^\rho_i$ and $\be_j=\bar{s}^\rho_j$.

Now, let $\te \in \Lp$. If $\te$ differs from $\al$ on two components, then since    $\al$ is 1-mistake away from $\bar{s}^p$, it follows from   Lemma \ref{IIfails} that $\te$ is also  1-mistake away from $\bar{s}^p$.  
If $\te$ differs from $\al$ on a single component $l\in N$, then there are three cases.

\noindent \textbf{Case 1:} Suppose that $l\in N\setminus\{i,j\}$.
Then, we have $\al_l=\be_l=\bar{s}^\rho_l\vartriangleright \te_l$, and by Lemma \ref{p_min2diff} part ii,   $\te_j\trianglerighteq \be_j=\bar{s}^\rho_j\vartriangleright\al_j$.\@ It follows that $\te$ differs from $\al$ on components $j$ and $l$, contradicting that  $\theta$ differs from $\al$ on a single component. 

\noindent \textbf{Case 2:} Suppose that $l=j$.
Then, since   $\te_k=\al_k=\bar{s}^\rho_k$ for each $k\in N\setminus \{j\}$, it directly follows that  $\te$ is   1-mistake away from $\bar{s}^p$.

\noindent \textbf{Case 3:} Suppose that $l=i$.  Then, consider $\ga$. Since $\displaystyle{\ga_{k^*}\vartriangleleft \be_{k^*}=\al_{k^*}}$,  by Lemma \ref{p_min2diff} part ii, for each $s\in  \{\be,\al\}$, it must be that $\ga_i\trianglerighteq s_i$ and $\ga_j\trianglerighteq s_j$.\@ Therefore, $\ga_i\trianglerighteq \al_i\vartriangleright\be_i$ and $\ga_j\trianglerighteq \be_j\vartriangleright\al_j$.\@ Then, since     $\al_i=\bar{s}^\rho_i$ and $\te_i\neq \al_i$, we have   $\ga_i\vartriangleright\te_i$.\@ Moreover,  since    $\te_j=\al_j$, we have  $\ga_j\vartriangleright\te_j$.\@  
   Therefore, $\te$ is dominated by $\ga$   on components $i$ and $j$, contradicting to  Lemma \ref{p_min2diff} part ii.
Thus, we conclude that $l\neq i$.
\end{proof}

\bigskip

\newpage

\newpage

\newpage

\newpage
\bibliographystyle{agsm}
\bibliography{choicematchingcombined}

\end{document}